# On the Orbit of Visual Binary ADS 8119 AB
## ($\alpha = 11^h\ 18^m\ 10^s.9$ and $\delta = +31^0\ 31'\ 44".9$)

**S.Siregar[1,2] and D. Hadi Nugroho[3]**
[1] Bosscha Observatory ITB, Lembang 40391
[2] Department of Astronomy, Bandung Institute of Technology, Indonesia
[3]Graduate Student in Astronomy

## Abstract

Xi Ursa Majoris ($\zeta$ Uma) historically is one of the most important double star in constellation Ursa Major, found by Sir William Herschel on May, 2, 1780 and the first binary successfully determined by using the principle of two body problem in 1828 by Savary. This star consists of two pair wide binary ADS8119 AB; in this case HD 98231(ADS 8119 A) as primary and HD 98230 (ADS 8119 B) as secondary. We have collected the observational data consist of separation angular ($\rho$) and position angle ($\theta$) from the observations in 1780 up to 2005 taken from Bosscha Observatory and other Observatories in the world. This paper presents the recent status of orbit binary system ADS 8119. By using Thiele Van den Bos method and empirical formula Strand's Mass-Luminosity relation we have determined the orbit and mass of ADS 8119AB. The result is;

### Orbital and Physical Element of ADS 8119 AB

| Dynamical Elements | Orbit Orientation | Masses-Parallax |
|---|---|---|
| $P = 60$ years | $i = 110^0.9$ | $M_1 = 1.4\ M_0$ |
| $e = 0.426$ | $\Omega = 104^0.7$ | $M_2 = 1.2\ M_0$ |
| $T = 1935.8$ | $\omega = 117^0.5$ | $p = 0".122$ |

## I. Introduction

Since the Bosscha Observatory ($\lambda=105^0$ and $\varphi=-6^030'$) was established in 1923 researches on visual binary stars played an important role in astronomical researches in Indonesia. Van Albada (1957) proposed a method to determine the orbit according to three base points of observation for moderate length of secondary trajectories. The ADS8119 AB is a visual double star, composed of HD 98231 as primary star with spectral type G0V and magnitudo $m_v$ = 4.41. The secondary star is HD 98230 with spectral type F8.5V and magnitudo $m_v$ = 4.87. Although the system is in northern hemisphere Voute an astronomer from this observatory has successfully measured the position angle and separation angle. There are many computational techniques proposed by astronomers. Algorithm developed gave us possibility to determine the elements orbit. In this work the Thiele-van den Bos method(vide; Alzner 2003) is used, that is employing three complete observations of the observing time, angular separation and position angle (t,$\rho$,$\theta$) together with the double area constant which must be obtained from additional data. This method already used by Siregar (1981) to determine orbit of visual binary COU 297. The observation data considered here were obtained between years 1780 up to 2005, collected from Bosscha Observatory and other sources. The figure 1 represents the curve of $\rho$ and $\theta$ as function of t

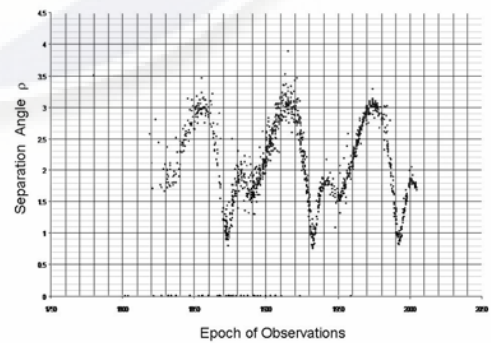

Fig. 1. Separation angle shown that ADS8119 AB has completed three revolutions

## II. Data and Notation

The data for this work collected from year 1780 up to 2004 are taken from catalogue of Washington Double Star (WDS). These data presented the star almost three time passed its periastron in average periode around 60 years. To determine the double areal constant C assumed the recent data more accurate then other, therefore for determining the orbit moreless 586 data of $(\rho, \theta)$ from epoch 1933.22 to 1992.42 are chosen. All of these data are shown in figure 1,figure 2, and figure 3

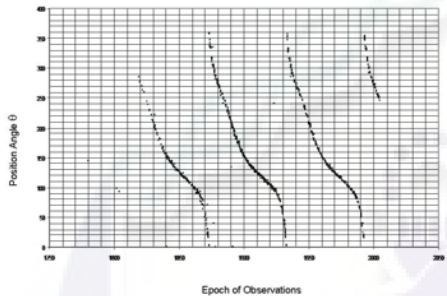

Fig. 2. Position Angle as function of observing time of ADS8119 AB from year 1750 up to 2004.The curve shown that the period of system is around 60 years

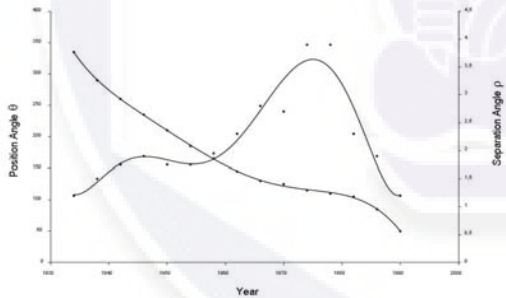

Fig. 3. The curve of corrected position angle and separation angle during period 1934 to 1990 are chosen for computing the Kepler constant

In this work we define some symbol such as;
$P,T,e,a,i,\omega,\Omega$ have their usual meaning for orbits
$\mu = 2\pi/P$ mean annual motion
$t_i$ = time at which the secondary star occupies its i-th position
$E_i$ = eccentric anomaly at time $t_i$
$t_{ij} = t_j - t_i$
$E_{ij} = E_j - E_i$
$V = E_{12}$ ; $U = E_{23}$ ; $W = E_{13}$
$(\theta_i, \rho_i)$ = coordinates of the secondary star at time $t_i$

$\Delta_{ij} = \rho_i \rho_j \operatorname{Sin}(\theta_j - \theta_i)$
C = the double areal constant of the apparent orbit

We define several supplement variables
$S_1 = (t_{23} - \Delta_{23}/C)$ ; $S_2 = (t_{13} - \Delta_{13}/C)$ ; $S_3 = (t_{12} - \Delta_{12}/C)$
$R = \Delta_{12}/\Delta_{13}$ ; $S = \Delta_{12}/\Delta_{23}$
$M = R(t_{12} - S t_{23})/(R t_{13} - S t_{23})$
$N = S(t_{12} - R t_{13})/(R t_{13} - S t_{23})$
$Q = (RS + R - S)^2 / 2RS$

## III. Fundamental Equations

If for the three pairs $(t_1, t_2)$, $(t_2, t_3)$ and $(t_1, t_3)$ the Thiele-Innes equations ( vide; Aitken, 1935) are writen in Kepler's equation, the result is;

$f_1 = U - \operatorname{Sin}U - \mu S_1$
$f_2 = V - \operatorname{Sin}V - \mu S_2$
$f_3 = W - \operatorname{Sin}W - \mu S_3$
$f_4 = U + W - V$

The three elliptical segments $S_1, S_2$ and $S_3$ must be positive. In case of an elliptical orbit, the equation (3.1) will lead to a solution for a mean annual motion only when ;

$$f = S_1^{1/3} - S_2^{1/3} + S_3^{1/3} > 0 \qquad (3.2)$$

It should be remembered that function f always increase with increasing C. Hence a slight increase of C may manage a sufficient increase of f, so that an elliptical solution may finally appear.

By choosing mean annual motion $\mu$ and regarding the condition U+W=V, the value of U,V and W are determined by iteration method of Newton (see for example, Press et.al, 1992). To start an iterative process, as initial value $\mu$ must be taken around a real mean annual motion. Arend (vide; Siregar 1988)) has shown that the best choice is to take;

$$\mu = 77.5\left[-f\left(S_1 - S_2 + S_3\right)\right]^{3/2} e^{-1.9138f} \qquad (3.3)$$

In case the convergence of the mean annual motion is not reached, i.e. its mean $V \neq U+V$, the cycle of calculation must be repeated with $\mu + \delta\mu$, $U + \delta U$, $V + \delta V$ and $W + \delta W$ where $\delta\mu$, $\delta U$, $\delta V$ and $\delta W$ are solution of linear systems;

$$\begin{bmatrix} -f_1 \\ -f_2 \\ -f_3 \\ -f_4 \end{bmatrix} = \begin{bmatrix} \dfrac{\partial f_1}{\partial U} & \dfrac{\partial f_1}{\partial V} & \dfrac{\partial f_1}{\partial W} & \dfrac{\partial f_1}{\partial \mu} \\[2mm] \dfrac{\partial f_2}{\partial U} & \dfrac{\partial f_2}{\partial V} & \dfrac{\partial f_2}{\partial W} & \dfrac{\partial f_2}{\partial \mu} \\[2mm] \dfrac{\partial f_3}{\partial U} & \dfrac{\partial f_3}{\partial V} & \dfrac{\partial f_3}{\partial W} & \dfrac{\partial f_3}{\partial \mu} \\[2mm] \dfrac{\partial f_4}{\partial U} & \dfrac{\partial f_4}{\partial V} & \dfrac{\partial f_4}{\partial W} & \dfrac{\partial f_4}{\partial \mu} \end{bmatrix} \begin{bmatrix} \delta U \\ \delta V \\ \delta W \\ \delta \mu \end{bmatrix} \quad (3.4)$$

This iterative process is stopped when; $\varepsilon^2 \leq \delta U^2 + \delta V^2 + \delta W^2 + \delta \mu^2$. In this work the tolerance factor $\varepsilon$ is less than $10^{-3}$ radian. Once the relationship between U,V and W has been obtained, the well-known expression;

$$e^2 = 1 + \frac{S(1 - CosU) + 1 - CosW - R(1 - CosV)}{Q} \quad (3.5)$$

define e as function of U,V and W. The behaviors of this function and various types of solutions have been published by Docabo(1985). For calculating the orbital parameters from U,V,W, $\mu$ and e is used the classical method such as;

$[U,V,W,\mu,e] \rightarrow$ [Thiele-Innes elements] $\rightarrow$ $[a,i,\omega,\Omega]$

Through mass-luminosity relation, primary and secondary masses are determined. For detail explanation see Alzner (2003). In case of an elliptical motion this mapping is guaranteed, but for non periodic orbits the Thiele-Innes elements must be modified (Dommanget 1983)

## IV. Application and Results

As an object study we shall consider of the binary ADS 8119 AB. The orbit for this double star has previously been calculated for Heintz (1967) and Mason et al. (1995). In order to make the example as comprehensive as possible, we shall start from three fundamental points and the apparent area constant C = 0.170 [radian][arcsecond]$^2$ /year, and standard deviation, $\sigma = 0.029$ [radian][arcsecond]$^2$ / year
The following table gives the three base points and the relative positions of the secondary star of ADS 8119 AB which will be used to determined the mean annual motion computed by this method is presented on the following table. A comparison is made with other results.

Table 1.Three base points of ADS 8119 AB which will be used to determine the mean annual motionThe orbital elements of visual binary ADS 8119 AB

| No | Year | θ | ρ | $t_{qp}$ | $\theta_{qp}$ | $\Delta_{qp}$ |
|----|------|-----|------|----|----|-------|
| 1 | 1950 | 210 | 1.75 | 10 | 50 | 2.965 |
| 2 | 1960 | 160 | 2.15 | 10 | 40 | 3.317 |
| 3 | 1970 | 120 | 2.6 | 20 | 90 | 4.320 |

Table 2. Orbital elements of ADS 8119 AB according to our computation and other author

| Elements orbit | Our works | Heintz (1967) | Mason et al. (1995) |
|----------------|-----------|---------------|---------------------|
| a["] | 2.401 | 2.53 | 2.536 |
| P[years] | 60 | 59.84 | 59.878 |
| T[year] | 1935.8 | 1935.17 | 1935.195 |
| e[.] | 0.426 | 0.414 | 0.398 |
| i[degree] | 110. 9 | 122.65 | 122.13 |
| ω[degree] | 117. 50 | 127.53 | 127.94 |
| Ω[degree] | 104.75 | 101.59 | 101.85 |

This system has masses $M_1$= 1.12 $M_0$, $M_2$ = 0.96 $M_0$ and dynamical parallax p= 0".122. Main while the Thiele-Innes constant follow;

A = 0".659 $\pm$ 0".128,
B = -1".378 $\pm$ 0".166,
F = 0".250 $\pm$ 0".079,
G = -3".453 $\pm$ 0".104.

## V. Acknowledgments

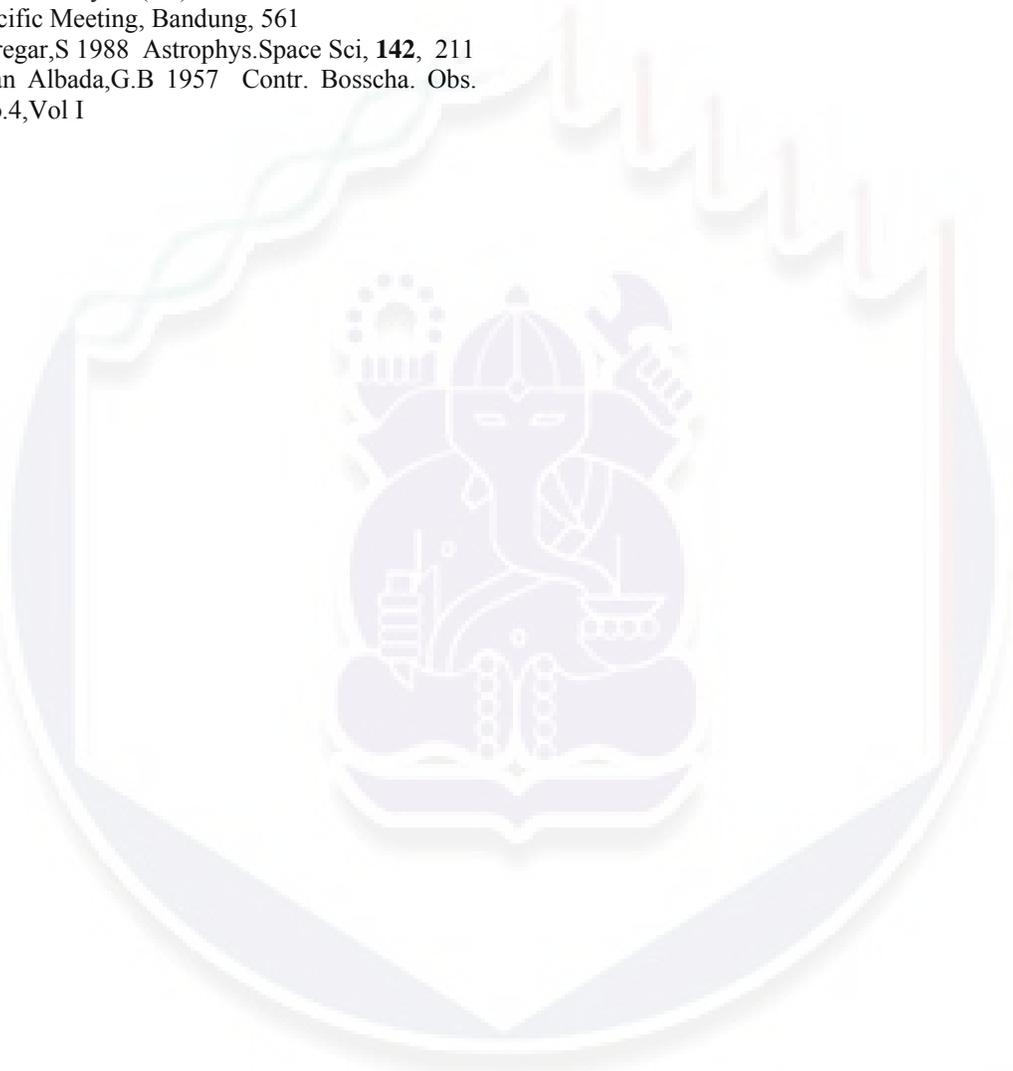